\documentclass[useAMS,usenatbib]{mn2e}
\usepackage{amsmath} 
\usepackage{subfigure}
\usepackage{times}
\usepackage{graphicx}
\usepackage{times}%
\usepackage{natbib}
\usepackage{times}%
\def\msun{{\rm M}_\odot}

\def\zsun{Z$_\odot$}

\def\simlt{\mathrel{\rlap{\lower 3pt\hbox{$\sim$}}\raise 2.0pt\hbox{$<$}}}
\def\simgt{\mathrel{\rlap{\lower 3pt\hbox{$\sim$}} \raise 2.0pt\hbox{$>$}}}

\title[The FMR of LGRBs hosts]
      {The metallicity properties of simulated long-GRB galaxy hosts and the Fundamental
      Metallicity Relation}
 \author[M.A. Campisi et al.]
        {M.A. Campisi$^1$\thanks{E-mail: campisi@dfm.uninsubria.it}, C. Tapparello$^2$, R. Salvaterra$^1$, F. Mannucci$^3$, M. Colpi$^2$\\
         $^1$Dipartimento di Fisica e Matematica, Universit{\`a} dell'Insubria, via
         Valleggio 7, 22100 Como, Italy\\
         $^2$ Dipartimento di Fisica G. Occhilalini, Universit\`a degli Studi di Milano-Bicocca, Piazza della
         Scienza 3, 20126 Milano, Italy\\
         $^3$INAF - Osservatorio Astrofisico di Arcetri, 
         Largo E. Fermi 5, I-50125, Firenze, Italy\\
}
\begin{document}

\date{Accepted ???. Received ???; in original form ??? }
\pagerange{\pageref{firstpage}--\pageref{lastpage}} 
\pubyear{2011}

\maketitle

\label{firstpage}


\begin{abstract}

We study the implication of the collapsar model for Long Gamma-Ray Bursts (LGRBs) on the metallicity properties of the host galaxies, by 
combining high-resolution N-body simulations with semi-analytic models of galaxy formation.
The cosmological model that we use reproduces the Fundamental Metallicity Relation recently discovered for the SDSS galaxies,
whereby the metallicity decreases with increasing Star Formation Rate for galaxies of a given stellar mass.  
We select host galaxies housing pockets of gas-particles, young and with different thresholds in metallicities, that can be
sites of LRGB events, according to the collapsar model. 
The simulated samples are compared with 18 observed LGRB hosts in the aim at discriminating whether the metallicity is a primary
parameter. 
We find that a threshold in metallicity for the LGRB progenitors, within the model galaxies, is not necessary in order to reproduce the observed 
distribution of host metallicities. The low metallicities of observed LGRB hosts is a consequence of the high star formation
environment.  The star formation rate appears to be the primary parameter to generate a burst event. 
Finally, we show that only a few LGRBs are observed in massive, highly extincted galaxies, while these galaxies are expected to produce many such events. We identify these missing events with the fraction of dark LGRBs.
\end{abstract}

\begin{keywords}
  gamma-rays: bursts -- host galaxies .
\end{keywords}

\section{Introduction}
\label{sec:intro}

The study of the host galaxies of Gamma-Ray Bursts (GRBs) is essential for understanding their nature. Current observations reveal that
long-duration GRBs occur in star-forming galaxies, in contrast to short-duration GRBs that are found in both early- and late-type 
galaxies. The star formation rate (SFR) in the short GRB host galaxies is often lower than 
in the LGRB's hosts (\citealp{ber06}, and references therein).
The difference in the observed host properties for short and long GRBs 
supports the idea that short and long GRBs have different progenitors.
Long-duration GRBs are believed to arise from the death of massive stars (collapsar model; \citealp{Conselice_etal_2005,fru06,tan07,wai07}, and references therein), most likely Wolf-Rayet stars, while short-duration GRBs are likely produced by the coalescence of compact objects in binaries 
\citep{li98,osh08}.
The discovery of the connection between long GRBs and core-collapse Type Ibc supernovae \citep[and references therein]{gal98,li06,woo06b} supports the collapsar model of long-duration GRBs \citep{mac99,mac01}. 
So far no supernovae have been found to be associated with short-duration GRBs.

In current core-collapse models for long-duration GRBs (LGRBs hereafter), young stars with initial mass $>30\,M_{\odot}$ should be able to create a black hole (BH) remnant. If the collapsing core has high angular momentum, the formation of the BH may be accompanied by a LGRB event (Woosley 1993; MacFadyen \& Woosley 1999).
The wind-driven mass loss in massive stars is a function of the metal content: high metallicity stars have 
strong stellar winds with main losses of both mass and angular momentum.
Instead, when metallicities are low ($0.1-0.3\,Z_{\odot}$; \citealp{woo06b}), the specific angular momentum of the progenitor star allows for the loss of the hydrogen envelope while preserving the helium core that can still carry rotation \citep*{woo06b,Fryer_Woosley_Hartmann_1999,Yoon_Langer_Norman_2006,yoon08}. 
The loss of the hydrogen envelope reduces the material that the jet needs to cross in order to escape, while the helium core
should be massive enough to collapse and power a LGRB.
\cite{vink05} explored LGRB models with even lower metallicites ($Z/Z_{\odot} \leq 10^{-3}$), showing that they have too low mass-loss rates to power a LGRB.
Thus, it may be possible to produce a LGRB in the metallicity range around $(0.1-0.3)$ of the solar metallicity.

The relationship between LGRB progenitors and their host environment has become matter of a hot debate, in recent years. In particular an open question concerns the metallicity of the LGRB host galaxies: is  the metallicity a discriminant for the formation of a LGRB? Do LGRBs form preferentially
in metal poorer galaxies? The observational information gathered so far indicates that most LGRBs are found in faint, star forming galaxies dominated by young stellar populations with sub-solar gas-phase metallicities, although there are a few host galaxies with higher metal content \citep[and references therein]{Prochaska_etal_2004,wol07,fyn06,price07,sav03,Savaglio_2006,sta08,levesque2010}.
We have to stress that low-metallicity progenitors do not necessary imply low-metallicity host galaxies. Indeed,
owing to the existence of {\it metallicity gradients} inside galaxies \citep{artale2011}, LGRBs could form from low-metallicity progenitor stars collapsing on metal deficient clouds  also in hosts with relatively high mean metallicities \citep{campisi09}. 

The existence of a possible metallicity discriminant in the GRB selected galaxies with respect to the overall field galaxy population can be tested comparing observed phenomenological relations in the two samples. In particular, the relation between stellar mass $M_*$ and metallicity $Z$ 
provides a good description of the properties of nearby galaxies \citep{Tremonti_etal_2004}. It has been shown that similar 
relations hold also at higher redshifts \citep{jabran2010,erb06} up to $z\sim 3-4$ \citep{mannucci09,maiolino08} with an evolution towards lower metallicities with increasing redshift. 

Recently, \cite{han2010} and \cite{levesque2010b} have compared the $M_*-Z$ relation for LGRB host galaxies with samples from the Sloan Digital Sky
Survey (SDSS) representative of the general star-forming galaxy population \citep{Tremonti_etal_2004}. They find that the metallicity 
content of low redshift LGRB hosts tends to fall off the $M_*-Z$ relation, and suggest that LGRBs do occur preferentially in lower metallicity galaxies. 
In order to further explore the origin of this offset, \cite{Mannucci2010c} compared the observed properties of LGRB hosts with those of field galaxies  
in light of the new Fundamental Metallicity Relation (FMR; \citealp{Mannucci2010a}). 
FMR is a tight relation between stellar mass $M_*$,  metallicity $Z$ and SFR.
Local SDSS galaxies define a surface in the 3D space of these three quantities, with metallicity well determined
by the stellar mass and SFR (with a scatter around this surface of about 0.05 dex; see \citealp{Mannucci2010a}). 
The key parameter that tighten the relation is a linear combination (in log scale) of stellar mass and  star formation rate:
 $\mu_{0.32}=\log M_*-0.32\,\log(\rm {SFR}).$
 \cite{Mannucci2010c} compared the FMR for galaxies with stellar masses down to $10^{8.3}\,\msun$ 
with the metallicity properties of 18 host galaxies of LGRBs with redshift $z<1$, 
for which $M_\star$, SFR and $Z$ were known. 
They found that while LGRB host galaxies show a systematic offset toward lower metallicities with respect to the $M_*-Z$ relation of field galaxies, 
no offset is present on the FMR.
This indicates that the deviation relative to $M_*-Z$ is due to the higher-than-average SFR observed in the hosts of the GRBs: the  lower metallicity content is a consequence of the occurrence of LGRBs in low-mass, actively star-forming galaxies which are, per se, metal poorer.  
In other words, the metallicity observed in LGRB hosts is exactly what is expected on the basis of their mass and SFR, with no apparent bias toward lower metallicities. 
In addition, \cite{Mannucci2010c} confirm that LGRB hosts have in general higher specific SFRs, i.e. higher SFR per unit stellar mass (SSFR hereon) as in \cite{savaglio2009}, suggesting that the condition for a galaxy to host a LGRB could be related to its ability to form stars in a 
efficient way.

Focus of this paper is at investigating whether the low metallicities of the progenitors to LGRBs requested by the collapsar model \citep{woo06b}
are in agreement with the observed properties of the galaxy hosts, in light of the findings by Mannucci et al. (2010).
The questions to answer are: do LGRBs preferentially select hosts with mean metallicities lower-than-average among star forming galaxies? Is the SSFR the primary physical parameter for the formation of a LGRB rather than the metal content? 
To this aim we use a simulated catalogue of galaxies, constructed by combining high-resolution N-body simulations with a semi-analytic prescription of galaxy formation. 
This allows us to identify different galaxies and to select those housing pockets of gas clouds with a metallicity threshold as requested by the collapsar model \cite{woo06b}.

The paper is organized as follows: in Section 2, we test the simulation outputs against the observed FMR and $M_*-Z$ relation, and show that the simulated galaxies provide a good description of the two relations.
Then in Section \ref{section3}, we analyze the properties of LGRB host galaxies, defined in the context of the collapsar model, 
and suitably selected within the simulated cosmic volume.
In Section 4, we explore the loci in the SSFR$-Z$ and $\mu_{0.32}-Z$ planes where we have the higher probability to host a LGRB.
Finally, in Section \ref{section5} we present our conclusions.


\section{The FMR of Simulated galaxy catalogues}
\label{section2}

\begin{figure*}
\centering
 {\includegraphics[width=8cm,angle=-90]{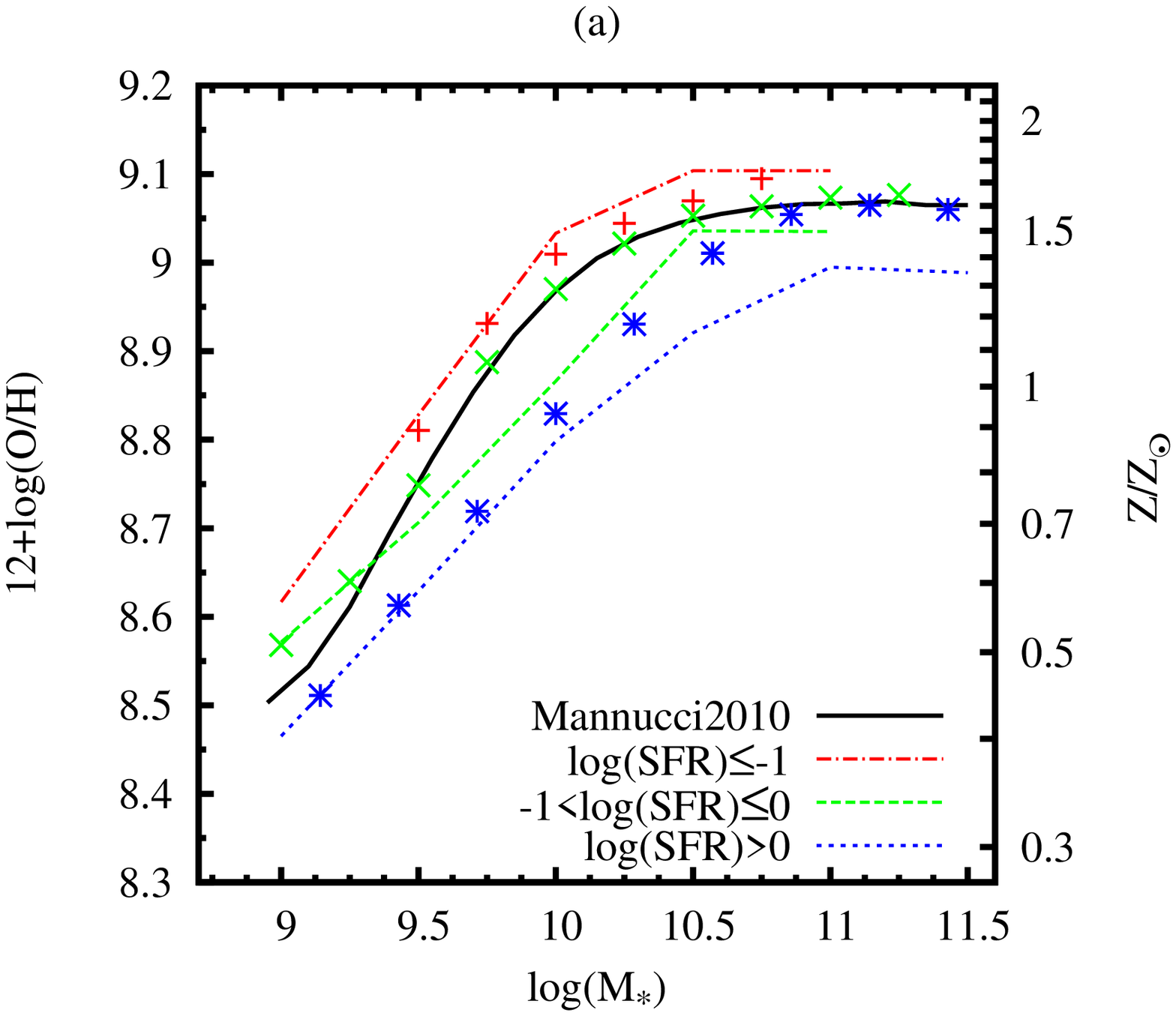}}\qquad
 {\includegraphics[width=8cm,angle=-90]{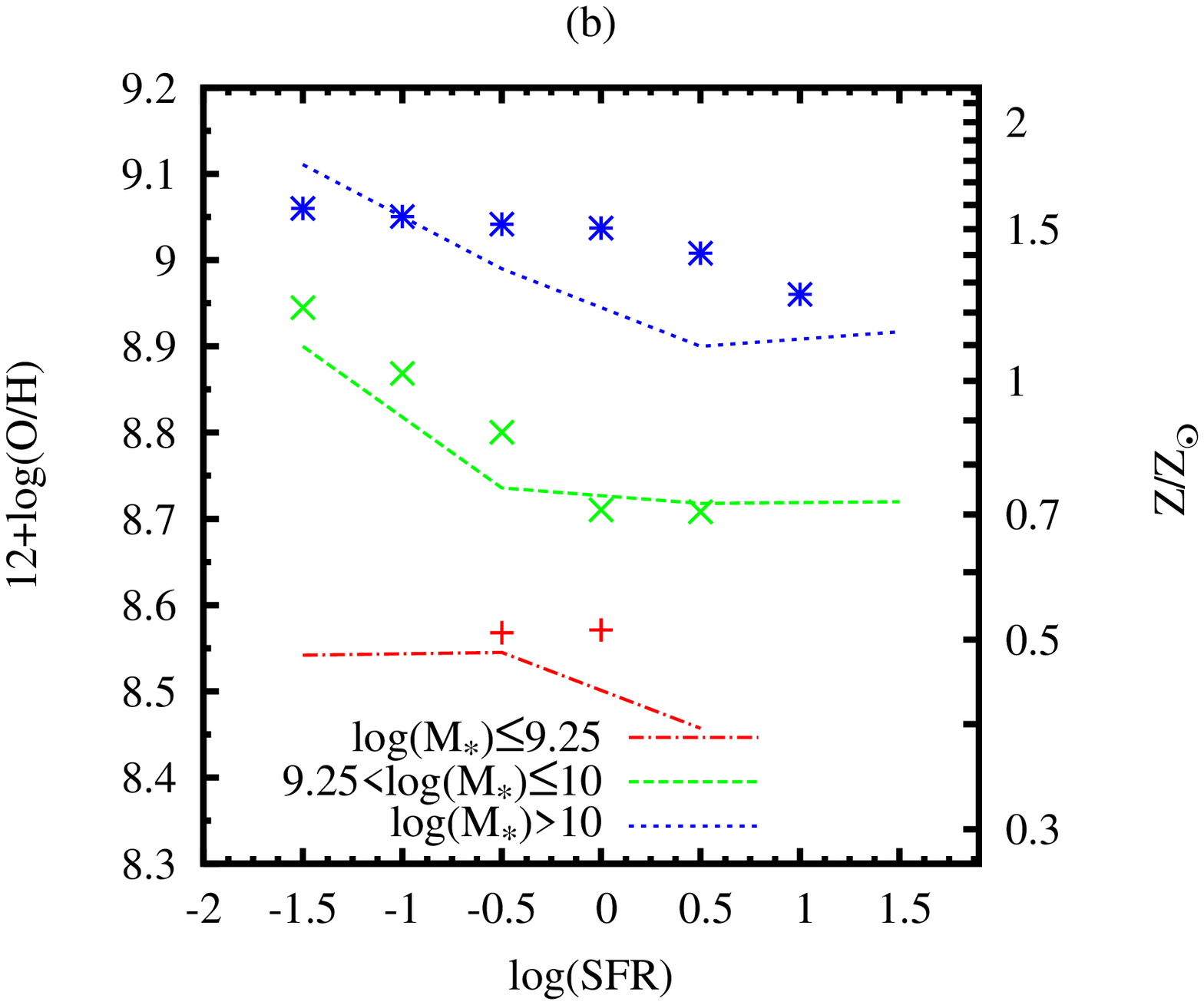}}
 {\includegraphics[width=8cm,angle=-90]{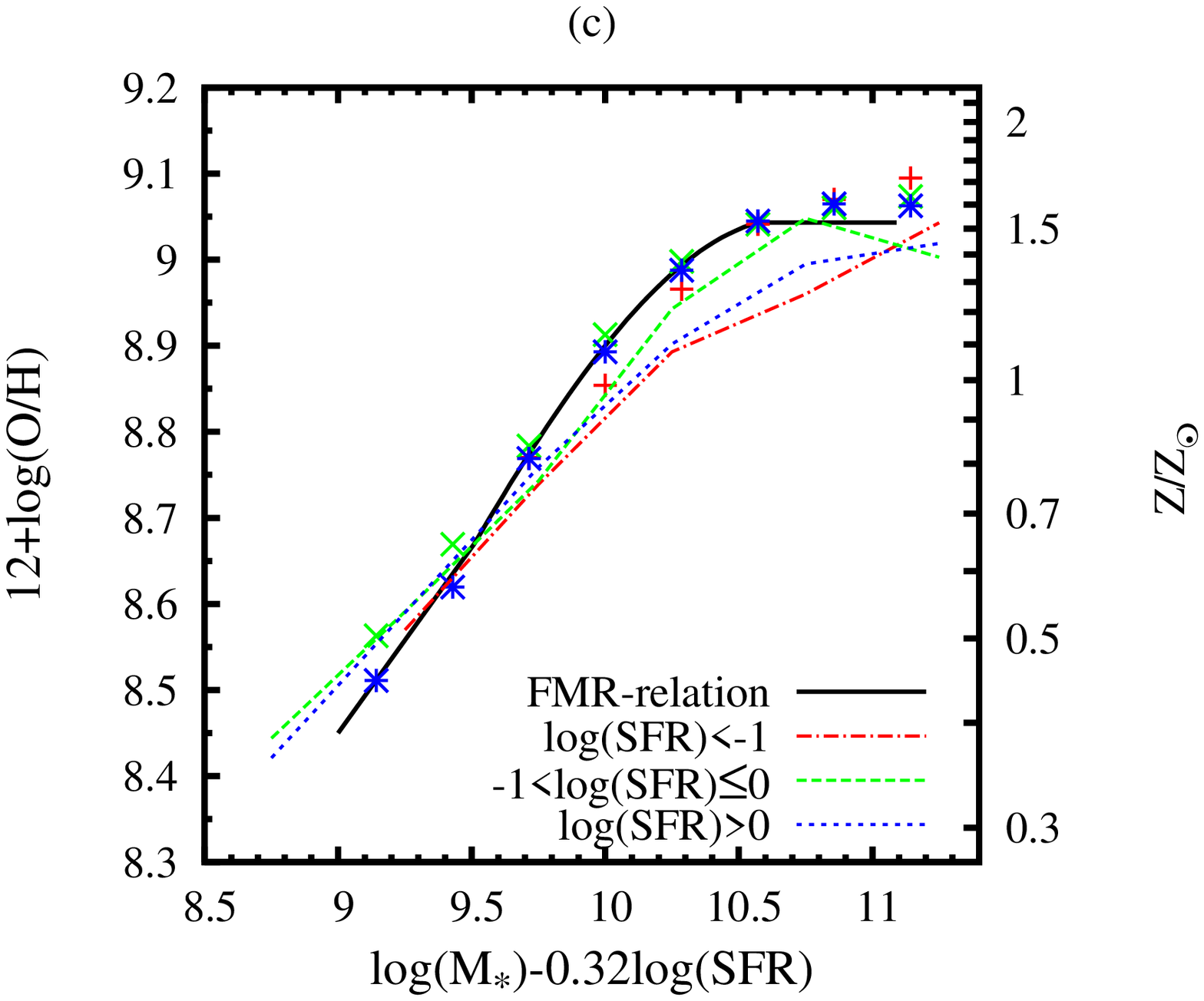}}
\caption{ Simulated galaxies in the local
 Universe ($0.07<z\leq0.3$) (lines) and galaxies of the SDSS (points in panels a,b,c); lines/points refer to the median values. Panel (a) shows the 
 metallicity $Z$ versus stellar mass $M_*$ (in units of $\msun$); colored lines/points show the $M_*-Z$ relation, in three selected SFR bins. Solid line is the observed relation (eq. 1 of the footnote).
Panel (b) shows the metallicity versus SFR (in units of $\msun$ yr$^{-1}$), for galaxies in
three different $M_*$ bins; 
Panel (c) shows the Fundamental Metallicity Relation, i.e. 
metallicity $Z$ versus $\mu_{0.32} \equiv\log(M_*) -0.32\log({\rm SFR})$ in three different SFR bins.
Dark-solid line refers to the  
FMR relation from Mannucci et al. (2011; eq. 2 of the footnote). 
}
\label{fig1}
\end{figure*}

In this Section we test whether the simulated galaxies reproduce the FMR.
Readers interested to the galaxy's catalogues  are referred to \citet{Wang_etal_2008}, \citet{Croton_etal_2006},
\citet{DeLucia_Blaizot_2007} and references therein for details on the
physical processes explicitly modelled. Here, we give a short description of the catalogue, 
focusing on the treatment of metal enrichment.

The galaxy catalogue used was constructed by \citet{Wang_etal_2008} with cosmological parameters consistent with the third-year WMAP results \citep{Spergel07}. The simulation corresponds to a box of $125\,h^{-1} {\rm Mpc}$ comoving length and a particle mass $7.78\times 10^8\,{\rm M}_{\odot}$. 
Simulation data were stored in 64 outputs and analysed with the post-processing software originally developed for the Millennium Simulation \citep{Springel_etal_2005}.
Merging history trees for self-bound structures extracted from the simulations were used as input for the Munich semi-analytic model of galaxy
 formation described in  \citet{DeLucia_Blaizot_2007}.
In particular, in the simulated galaxies the metals are produced primarily in the supernovae and are deposited directly in the cold gas present in the disc of the galaxy (istantaneous recycling approximation; \citealp{Croton_etal_2006,DeLucia_Kauffmann_White_2004}). \footnote{The interpretation of the FMR in the context of the simulated catalogues is beyond the goal of present work.}

We limit our analysis to galaxies with stellar mass larger than $8\times 10^8\,{\rm M}_{\odot}$, which is above the resolution
limit of the N-body simulations used.
In order to be consistent with the observed local sample of SDSS galaxies, we select, in our simulated cosmological volume, galaxies with 
redshift $0.07< z \leq0.3$ as in \cite{Mannucci2010a}, for which the SFR, stellar mass and mean metallicity are known.
The resulting sample includes $1075878$ galaxies.

Fig.\ref{fig1} shows the result of our comparison: panel (a) refers to the $M_*-Z$ plane where the metal content $Z$ is express in terms of the oxygen to hydrogen abundance ratio $Z=12 +\log({\rm O/H})$ \footnote {The best fit to the observed $M_*-Z$ relation for galaxies in the local universe is \citep{Mannucci2010a}: 
\begin{equation}
12+\log ({\rm O/H})=8.96 + 0.31 m -0.23 m^2-0.017m^3+0.046m^4
\end{equation}
where $m=[\log(M_*/ \msun)-10]$. We will refer to eq. 1 hereon as $M_*-Z$ relation.
} (left $y-$axis), and in solar units (right $y-$axis); panel (b) refers to the SFR$-Z$ plane; and in panel (c) refers to the $\mu_{0.32}-Z$ plane where
the FMR is defined \footnote {The analytical expression of the FMR for
local galaxies is \citep{Mannucci2010c}:
\begin{equation}
\begin{array}{rl}
12+\log({\rm O/H})&=8.90+0.37m-0.14s-0.19 \, m^2                    \\
           &+0.12ms-0.054\,s^2~~~~~~~~~~\rm{for}~\mu_{0.32}\ge9.5 \\
           &=8.93+0.51(\mu_{0.32}-10)~~\rm{for}~\mu_{0.32}<9.5 \\
\end{array}
\end{equation}
\noindent 
where $m=[\log(M_*/\msun)-10],$ and $s=\log ({\rm SFR})$ with SFR expressed in units of $\msun\,{\rm yr}^{-1}$.}; 
In the first three panels, lines refer to the simulated galaxies and points to SDSS galaxies.
In more detail, panel (a) shows the median metallicity for the simulated galaxies in different SFR bins
as a function of the stellar mass. On average each bin contains more than $10^3$ galaxies. 
Galaxies with higher SFRs show systematically lower metallicities at a fixed $M_\star$. Indeed this trend is clearer in panel (b) where the SFR$-Z$
relation is plotted in three different $M_\star$ bins. 
This finding is in agreement with the behavior of the observed SDSS galaxies in the same redshift range (depicted as points).
As shown in panel (c), once the dependency on the SFR and $M_*$ is included into the parameter $\mu_{0.32}$, the medians, in each 
SFR bin, collapse into a single, well defined line consistent with the observed FMR shown as dark-solid line. 

In conclusion our tests have shown that the simulated galaxy population 
provides a good description of the observed properties of local SDSS field galaxies.
\begin{figure*}
\centering
{
\includegraphics[width=8cm] {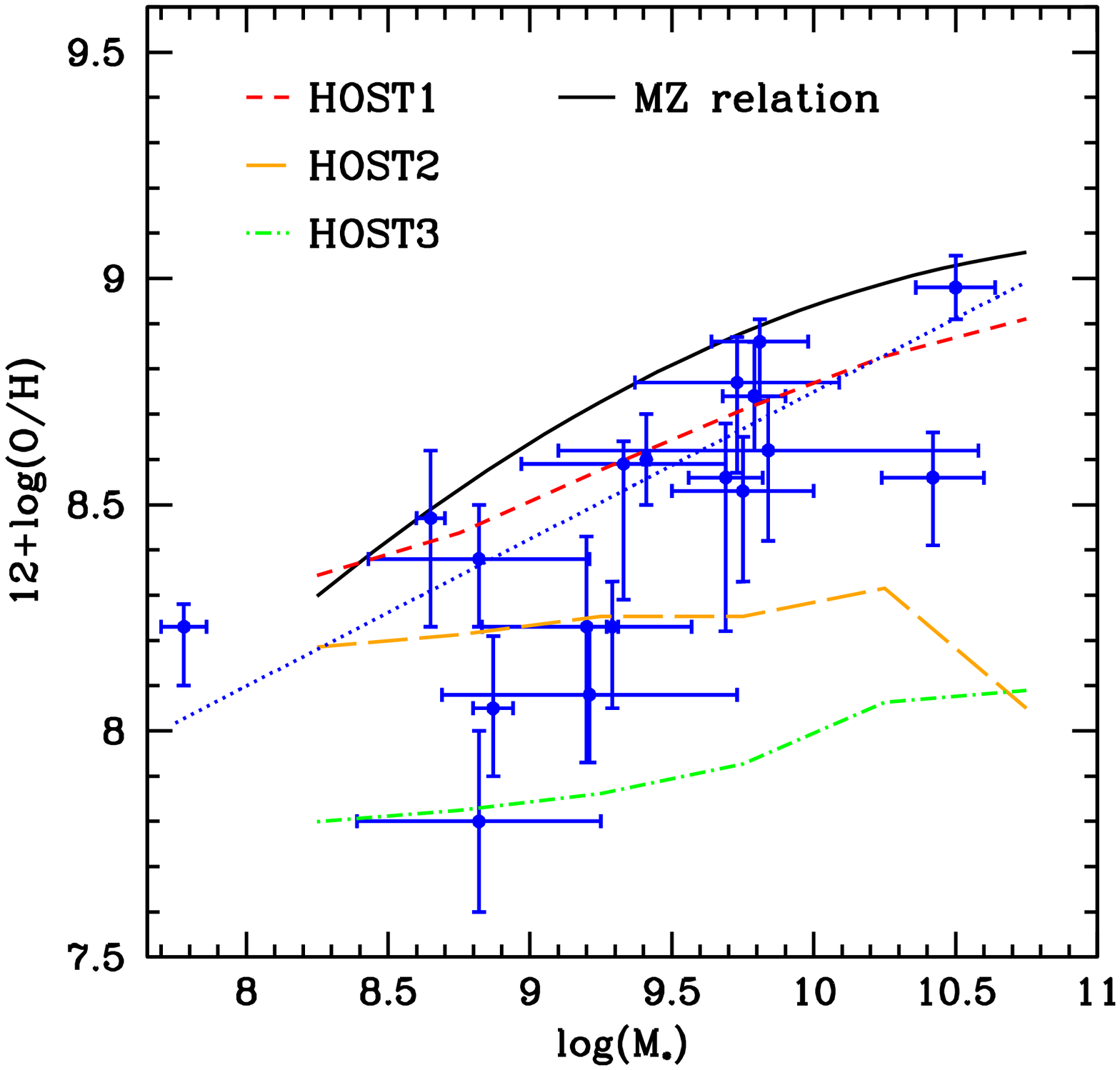}
\includegraphics[width=8cm] {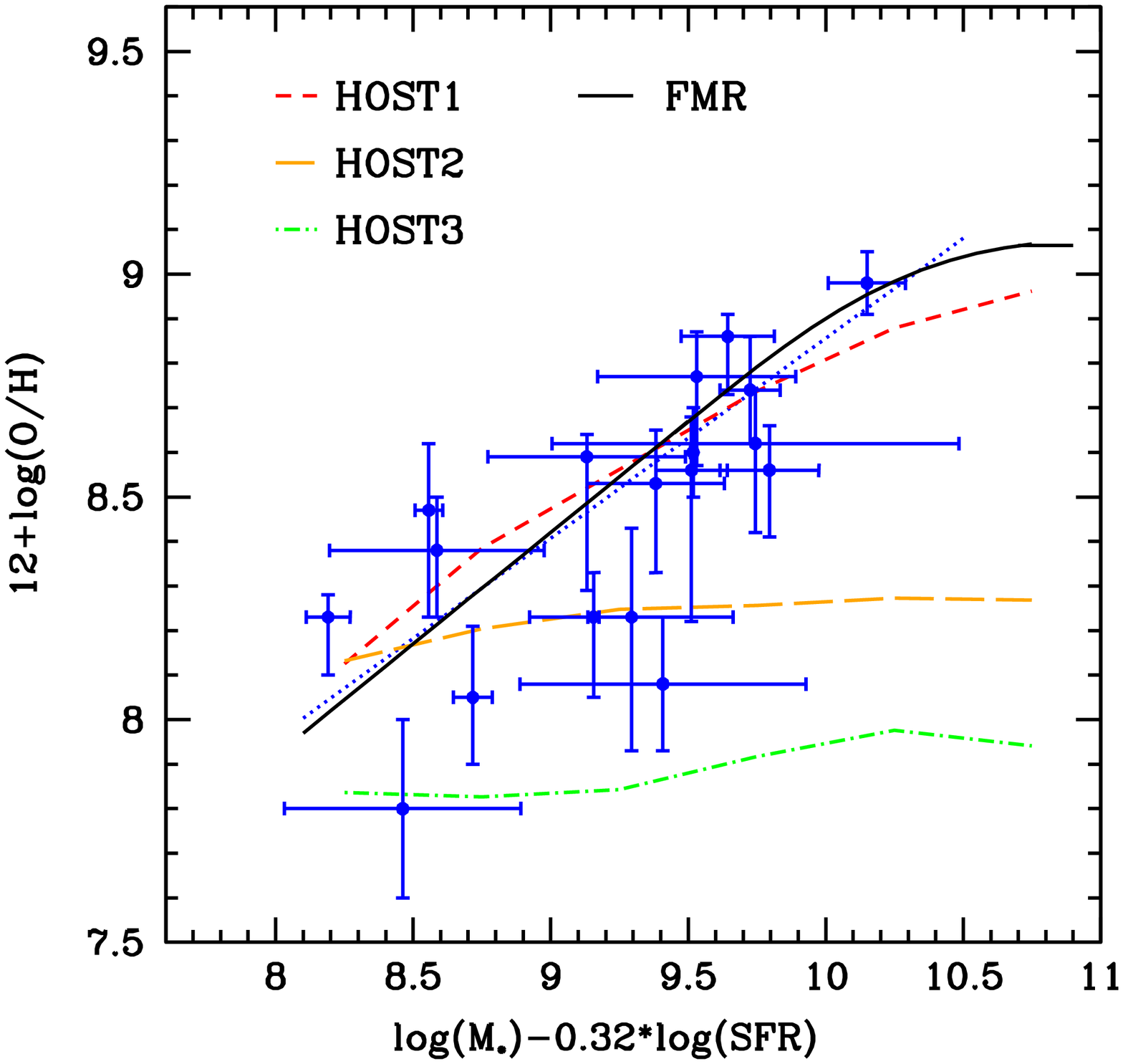}
}
\caption{{\it Left}: Mass-Metallicity relation for host simulated galaxies. 
 The solid-black line is the Tremonti relation shifted to higher redshifts (Jabran et al.2010), 
the red-dashed line is the median for HOST1, the yellow-long-dashed and green-dot-dashed line refer
to HOST2 and HOST3, respectively.
Points are the observed host galaxies of our LGRB sample and the blue-dotted line is their best fit.
{\it Right}: FMR  for host simulated galaxies. We use the same line coding of the right plot. 
Black-solid line the FMR Mannucci et al. 2010a.
}
\label{fig5}
\end{figure*}

\section{Simulated LGRB host galaxies}
\label{section3}
In this Section, we use the simulated galaxy sample to select possible LGRB hosts in the light of the collapsar scenario.

Following \cite{woo06b}, LGRBs occur more favorably in stellar environments with metallicities lower than 0.1-0.3 $Z_\odot$. 
Accordingly, we extracted from the simulated catalogue the galaxies which contain 
{\it pockets} of gas ``particles''  that can be sites to progenitors to LGRB events.
Pockets have a metallicity $Z_{\rm prog}$ and age $t_{\rm age}$, and the host has metallicity $Z_{\rm host}$ computed as mean of the metallicity of all pockets present inside the simulated galaxy. 
According to  \cite{campisi09}, we create three host's catalogues
with different properties:

\smallskip

\noindent
(i) HOST1, obtained by selecting galaxies containing stars with
  age $t_{\rm age}< t_*= 5\times 10^7 {\rm yr}$;

\noindent
(ii) HOST2, including galaxies with stars of age $t_{\rm age}< t_*$ and
  metallicity $Z_{\rm prog}\leq 0.3\, Z_{\odot}$;
  
\noindent
(iii) HOST3, defined by selecting galaxies containing stars with age $t_{\rm age}<t_*$ 
  and metallicity $Z_{\rm prog}\leq0.1 \, Z_{\odot}$.

\smallskip

We computed the number of stars ending their lives as LGRBs, assuming a Salpeter\footnote{We check that an alternative IMF, 
as the Chabrier, does not change the selection of host galaxies in our simulation.}
Initial Mass Function (IMF), and on average (over all cosmic times) one LGRB event every 1000 supernovae \citep{Porciani_Madau_2001,campisi2011}.
The number of expected LGRBs in the $k$-th galaxy can be computed as $N_{k,i}\propto\,\,\zeta_{{\rm BH},i}\,\,M_{*,i,k}(z)$, where $\zeta_{{\rm BH},i}$ is the fraction of stellar mass that will produce black holes in case $i$ (where $i$ runs over the three catalogues, $i=1,2,3$) and $M_{*,i,k}$
is the stellar mass in the $k$-th galaxy that satisfies our selection criteria for the $i$-subsample (see Sec.3 in \citealp{campisi09} for details).
We point out that we select host galaxy candidates assuming constraints on the metallicity of LGRB progenitor stars, not on the overall metallicity $Z_{\rm host}$ of the host galaxy. 

\subsection{$M_*-Z$ and FMR of simulated LGRB hosts}

In this Section, we focus on the position of LGRB hosts in the planes shown in Fig.1 and on the comparison with the observed LGRB hosts. 

We first include the simulated galaxies belonging to the three samples HOST1-2-3 in the $M_*-Z$ and $\mu_{0.32}-Z$ planes.
The metallicity, in this case, refers to the galaxy as a whole, and is extracted as $Z_{\rm host}$ from the catalogue. For each sample, we computed the median metallicity values as a function of the stellar mass $M_*$ and similarly, for FMR, as a function of $\mu_{0.32}$. 
In each sample, the median is obtained by weighting the host galaxies with  the rate of occurrence of LGRBs  in their pockets.
In addition we restrict to galaxies at $z\sim 0.5$ being the mean redshift of the observed sample of LGRB hosts.  

Fig. \ref{fig5} shows the results for HOST1-2-3 and the comparison with the LGRB's data points. 
The left (right) panel refers to the $M_*-Z$ relation (FMR), and lines refer to the median computed for each sample as described in the figure caption.
The black line in the right (left) panel refers to the $M_*-Z$ relation observed for galaxies at $z\sim 0.8$ \citealp{jabran2010} (to FMR as in \citealp{Mannucci2010c}). 
The blue data points and blue dotted-line refer to the observed sample of 18 LGRB hosts. 

In the left panel of Fig. 2, HOST1-2-3 show a systematic offset toward lower metallicities compared to the observed $M_*-Z$ relation for field galaxies.
HOST2 and HOST3, corresponding to low cut-off in $Z_{\rm prog}$, display a flat $M_*-Z$ relation.
In these two samples, single simulated LGRB host galaxies with large $M_*$ may have high metallicities (i.e. high $Z_{\rm host}$) as illustrated and discussed in the Appendix A. However, their contribution to the median, weighted for the rate of LGRB housed
in the galaxy, is negligible as they are found to be site of only few LGRB events. 
In the right panel of Fig. 2, HOST1-2-3 are compared with the FMR.
HOST1 reproduces the observed relation for field galaxies quite well. 
By contrast HOST2 and HOST3 show significant deviations from the FMR, as in the $M_*-Z$ plane.

The data of the 18 LGRB hosts are  plotted in Fig. 2, and the blue-dotted line describes the best linear fit to the data (blue-points with associated
errors are as reported in \cite{Mannucci2010c}).  
In the $M_*-Z$ plane, GRB060218 is included in the fit  though it  lies at $\log (M_*/\msun)=7.78$ which is outside the range of
mass accessible from the simulated volume.

In both planes, HOST1 is the sample providing the best description of the available data.
In particular, this sample reproduces well the FMR for the observed LGRB hosts and the systematic offset in the $M_*-Z$ relation.
This is due to the fact that in our simulation the probability to host a LGRB is higher in galaxies
with higher SFRs.  

To further investigate these findings, we consider the FMR of the simulated 
catalogues dividing galaxies in subsamples with SFRs in different bins.  The results are shown in the left panels 
of Fig.\ref{fig2} of Appendix A.
We find that the FMR is well reproduced again by all subsamples of the HOST1 group,
and by the subsample of galaxies with high SFR ($\log ({\rm SFR})>0.9)$ of the HOST2 group. 
By contrast, none of the HOST3 subsamples reproduce the data since the constraint on metallicity in the progenitors
$Z_{\rm prog}$ implies also a severe cut in the metallicity of the host $Z_{\rm host}$.
Therefore, we can conclude that 
collapsar model predicting a strong metallicity bias
can not be easily reconciled with the observational evidence that LGRB hosts do follow the FMR.

\section{Probability distributions and dark LGRBs}

It has been noted that observed LGRBs tend to occur in galaxies with high specific star formation rates (SSFRs)
\citep{savaglio2009,Mannucci2010c}. The SSFR  or its inverse, the so-called doubling time, are good 
markers of the ability of a galaxy to form stars in an efficient way. \cite{Mannucci2010c}
have shown that the observed LGRB hosts have always doubling times shorter than the Hubble time at the redshift of the object. 
This holds true for additional 17 LGRB hosts at $z<1$ of unknown metallicity  for which stellar mass $M_*$ and SFR have 
been measured \footnote{Data are from the GHOST database (www.grbhosts.org)}. 
We note that none of them has $\log({\rm SSFR})<-9.95$ (with SSFR in units of yr$^{-1}$) 
confirming that all observed LGRB hosts have high SSFRs and doubling times shorter
than the Hubble time, at the redshift of the burst.  The position of LGRB
hosts in the SSFR-$Z$ plane may thus provide new hints on the host
galaxy population.

\begin{figure}
\centering
{
\includegraphics[width=8cm, angle=-90] {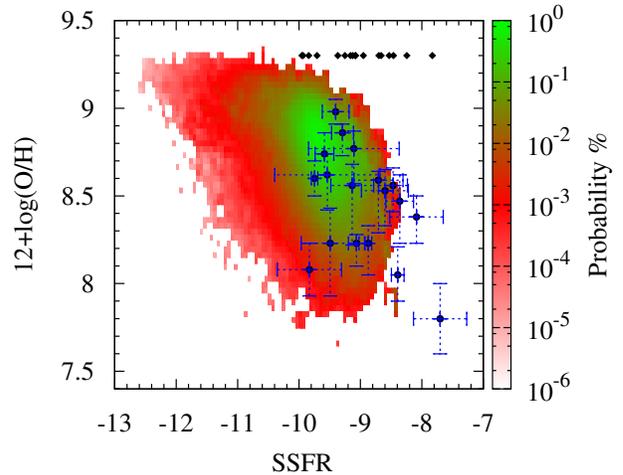}
}
\caption{
SSFR (in units of yr$^{-1}$) versus  metallicity $Z$. Blue points are the data with associated error bars
for the observed sample of 18 LGRB hosts \citealp{Mannucci2010c}. 
Black point at the top of the figure refer to 17 LGRBs for which the metallicity is not measured yet. 
The shaded area gives the map of the weighted probability to observe a GRB event for HOST1 sample. 
The lack of simulated galaxies with $\log {\rm SSFR}>-8.5$ is due to the resolution limit of the simulation used.
 }
\label{fig6}
\end{figure} 

In Fig.~3 we collect all observed data (blue dots) of the 18 LGRB in the SSFR-$Z$ plane 
(the 17 SSFRs for the additional hosts with lack of Z measurement are plotted as black point at the top
of the figure). In the same plane, the color-coded area gives the normalized probability map to have a LGRB event in galaxies of 
the HOST1 sample (see Appendix for our analysis on HOST1-2-3 in separated bins of SFR or mass $M_*$).
The probability is computed normalizing the number of LGRB events of each galaxy for the total number of events. 
Most but not all observed LGRB hosts fall in the green zone where the expected probability for the HOST1 sample is the highest. The simulated galaxies cover a wide area in the SSFR-$Z$ plane and extend also to low SSFRs.
About 15\% of all simulated LGRBs reside in galaxies with $\log(\rm {SSFR})<-9.93$, i.e. with doubling times longer than the Hubble time at
$z=0.5$. Therefore the model does not exclude hosts with low SSFR and we may still lack of LGRBs due to the low statistics related to the bias
against identifying galaxies with low SFRs, or to dust extinction (see below).

\begin{figure*}
\centering
{
\includegraphics[width=8cm, angle=-90] {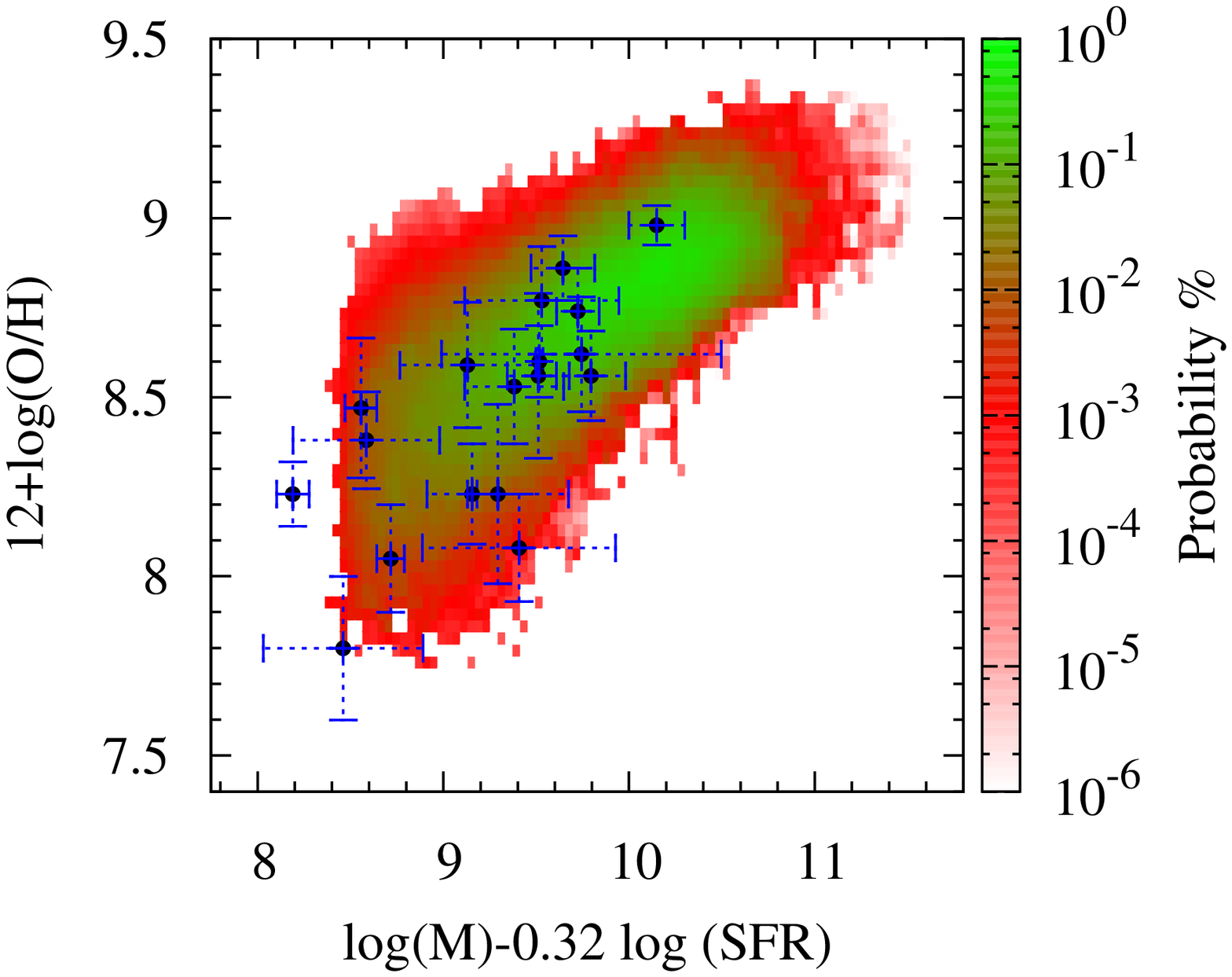}
\includegraphics[width=8cm, angle=-90] {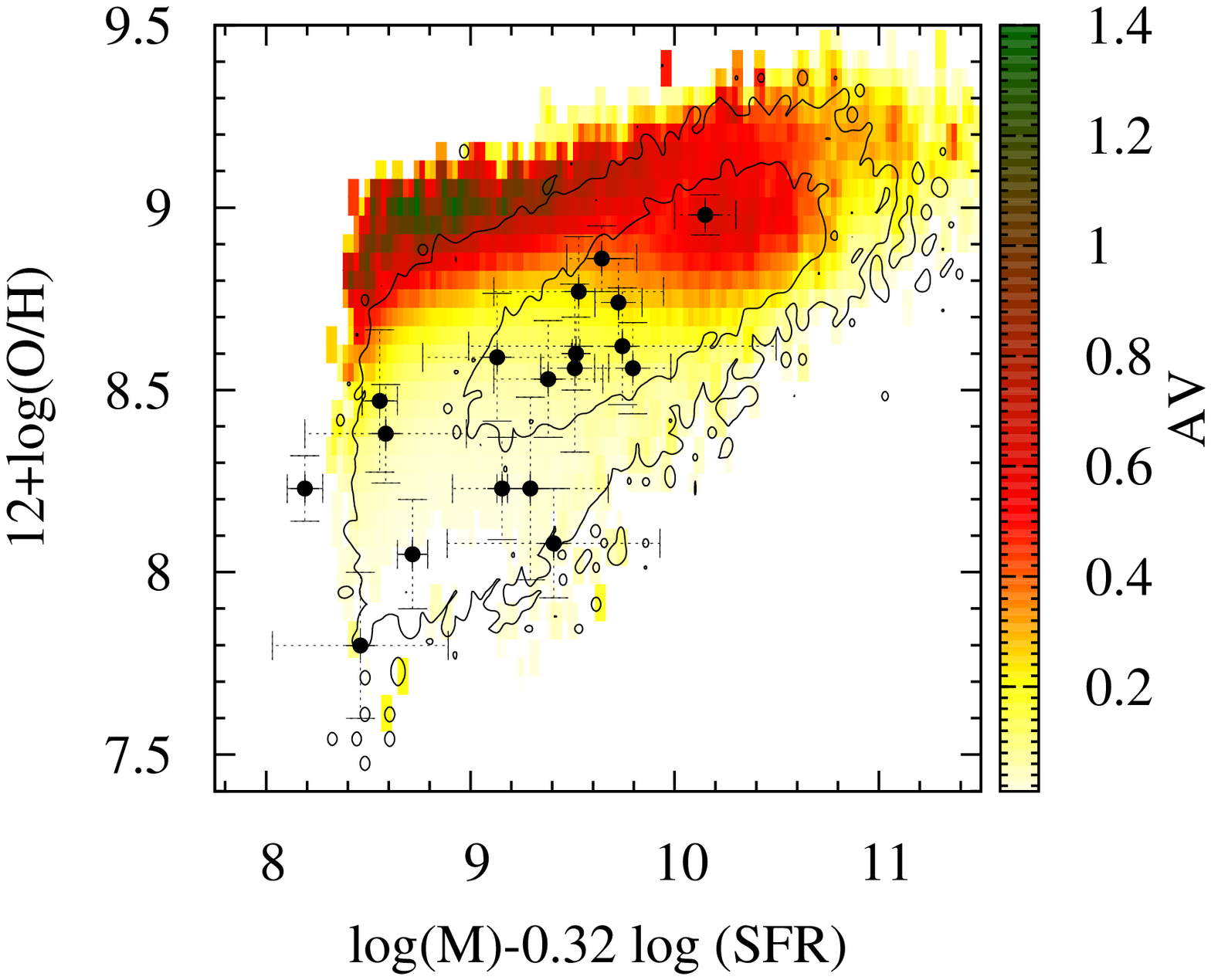}

}
\caption{$\mu_{0.32}-Z$ plane. In the left panel we show the color coded
probability map of housing a LGRB for galaxies of HOST1 (with $z\sim 0.5$).
Blue data points refers as in Fig. 3 to the observed LGRB hosts.
In the right panel,  we show the color coded probability map for a galaxy to
have a given averaged extinction coefficient $A_v\rm{[mag]}$
Simulated host galaxies are selected among the HOST1 catalogue at $z\sim 0.5$. The contours 
correspond to the 50\%(smaller area) and the 99\% (bigger area) of the entire population.}
\label{figAv}
\end{figure*}

Similarly, in the left panel of Fig.~4 we derive the color coded probability map of having a LGRB in a galaxy  (with $z\sim 0.5$) 
of the HOST1 sample with given metallicity and $\mu_{0.32}$ (the lack of simulated galaxies with $\mu_{0.32}<8.2$ is due to the resolution limit of the N-body simulation used). Data-points are also shown for the observed LGRB host galaxies. 
As in Fig.~3, most of the simulated hosts cluster in a very
well defined region of the FMR, and these loci are populated by the observed data.

We note the existence of a region (i.e., the green area in the interval  $10<\mu_{0.32}< 10.5$ 
corresponding to $8.6<12+\log({\rm O/H})<9.1$) 
with high values of probability of LGRB occurrence that is underpopulated of 
observed LGRB hosts.  The lack of observed 
hosts  may be linked to the existence of  dark
LGRBs, i.e. LGRBs with bright X-ray afterglow that are faint or not detected
in optical. It has been shown \citep{perley2009} that most of these objects 
can not be detected at optical wavelengths being strongly absorbed by dust
in their host galaxies. Here, we argue that dark LGRBs may reside in
hosts with high metal content (and therefore more dusty) that could populate
the empty region in the $\mu_{0.32}-Z$ plane. Indeed the only point present in this
region is relative to the host galaxy of GRB 020819, that is classified as a 
dark burst \citep{levan03,klose03,yoldas2010}.

We briefly analyze the simulated galaxies (HOST1) with high probability to host a LGRB 
in the empty region at high $\mu_{0.32}$, pointing out that they have higher stellar mass ($10<\log (M_*/\msun)<11$), 
and they show higher extinctions.
In the right panel of Fig.~4 we show the color-coded map
distribution of the total extinction coefficient $A_v$ in the V band (Buser V filter).
The contours contain the 50\% (smaller area) and the 99\% (bigger area) of the entire population of
simulated hosts showed in the left panel of Fig.~\ref{figAv}.
Again, the blue data points are the observed host galaxies. 
The principal result of this figure can be summarized as follows: despite 
the higher probability for galaxies with high $\mu_{0.32}$
to house a LGRB, their higher extinction prevents us from observing the optical afterglows of their LGRBs, instead
the galaxies with lower extinction, and thus lower metallicity, are favorable to be observed.
The white and yellow  regions in the map refer to areas of low extinction where the bulk of the observed LGRBs are found.
Similarly in the observation of core-collapse SNe, the missing in the Optical and IR observation
is supposed to be correlated with the extinction of dust in the host
galaxy \cite{maiolino2002,mannucci2003,cresci2007}.

However, we notice here 
that the present sample of observed LGRB hosts is not a complete or well 
controlled sample, and several selection effects could exist. In particular,
we can not use the present sample to investigate the relative fraction of 
LGRB hosts detected as function of $\mu_{0.32}$
Thanks to the unbiased simulated galaxy catalogues, we are able to predict the distribution on the FMR plane of
host galaxies, waiting for future observations of dark GRB host galaxies.

\section{Conclusions}
\label{section5}

In this paper, we studied the properties of LGRB host galaxies comparing 18 observed hosts with a catalogue
of simulated galaxies constructed combining high-resolution N-body simulations
with an up-to-date semi-analytic model of galaxy formation.
The simulated catalogue reproduces the offset on the observed $M_*-Z$ relation 
and the tighter Fundamental Metallicity Relation (FMR)  of the field galaxy populations (in the redshift interval $0.07<z\leq0.3$) 
with no need of introducing new free parameters.

Aim of our investigation was at studying the relative importance of metallicity and SFR in the characterization of a LGRB and of its host. To this purpose we studied the effects imposed by the presence of a metallicity threshold in the generation of a LGRB on the host galaxies.
We extracted, from the main simulated catalogue,
three  samples of  LGRB hosts, housing different pockets of gas particles 
with different metallicities $Z_{\rm prog}$ that could be potential sites to LGRB formation.
The first sample, denoted as HOST1, comprises galaxies undergoing star formation but with no threshold on
the metallicity ($Z_{\rm prog}$).
HOST2 comprises instead  star forming galaxies with lower metal content, i.e. a cut at 
$Z_{\rm prog}=0.3$~\zsun, and  HOST3 at $Z_{\rm prog}=0.1$~\zsun.

Our analysis shows that HOST1 is the sample providing a very good description of the currently available LGRB host 
dataset. In particular, the simulated sample is able to reproduce, at the same
time, the systematic offset of LGRB hosts toward lower metallicities in 
the $M_*-Z$ relation, and the tightness of the FMR in the $\mu_{0.32}-Z$ plane.
In the simulated sample, the probability of housing a 
LGRB is higher in active star-forming galaxies. As a consequence, when weighted for this 
probability the simulated hosts tend to show lower metallicities at fixed 
stellar mass, in the $M_*-Z$ plane. The HOST1 sample also reproduces 
the general trends in the SSFR$-Z$ plane of observed LGRB hosts. 
By contrast, HOST2-3 predict too lower metallicities in the
$M_*-Z$ relation with respect to what observed, and fail to reproduce the FMR of the available LGRB host sample. Also
they exhibit an inverted behavior in the SSFR$-Z$ plane with respect to the
observed LGRB hosts. 

The close match of HOST1 with both the LGRB dataset and the FMR observed in 
the field, indicates that the request of a cutoff in the metallicity of the progenitor stars
to LGRBs is not compelling. The low metallicity that is observed in the hosts
of LGRBs appear to be a consequence of the higher probability of producing a LGRB in
galaxies of low mass that have a higher star formation rate. Since galaxies of a given stellar
mass that are more star-forming are metal poorer, our analysis suggests the SFR  
to be the primary marker of LGRB formation.  

The observed sample of LGRB hosts is not complete and could be affected by strong bias effects.
Indeed, the sample comprises mainly LGRBs that displayed an optical afterglow.  There exists however,
LGRBs with X-ray afterglows that do not show any optical counterpart, i.e. the so called
{\it dark} LGRBs. Many of these are found in dust rich environments (Perley et al. 2009).
It is unclear whether dust is spread over the galaxy or confined to the LGRB local
environment. In the first case the host should have a high mean metallicity. 
These LGRBs should populate the loci of high metallicity of the FMR that are
devoid of objects, at present. The lack of
such hosts is suggestive that the optical afterglow of such  LGRBs in these
galaxies suffer sever extinction, and so we are led to identify these missing events 
with the fraction of dark LGRB, according to our probability analysis study.

\section*{Acknowledgements}
We are indebted to Dr. Jie Wang and Dr. Gabriella De Lucia for making available their simulated galaxy
catalogues and simulation outputs, and for useful comments. We thank Sandra Savaglio for having collected the data on GRB hosts.\\
\bibliographystyle{mn2e}
\bibliography{HostFMR2}

\begin{appendix}
\section{Details on the hosts' samples }
\begin{figure*}
\centering
{
\includegraphics[width=6.5cm] {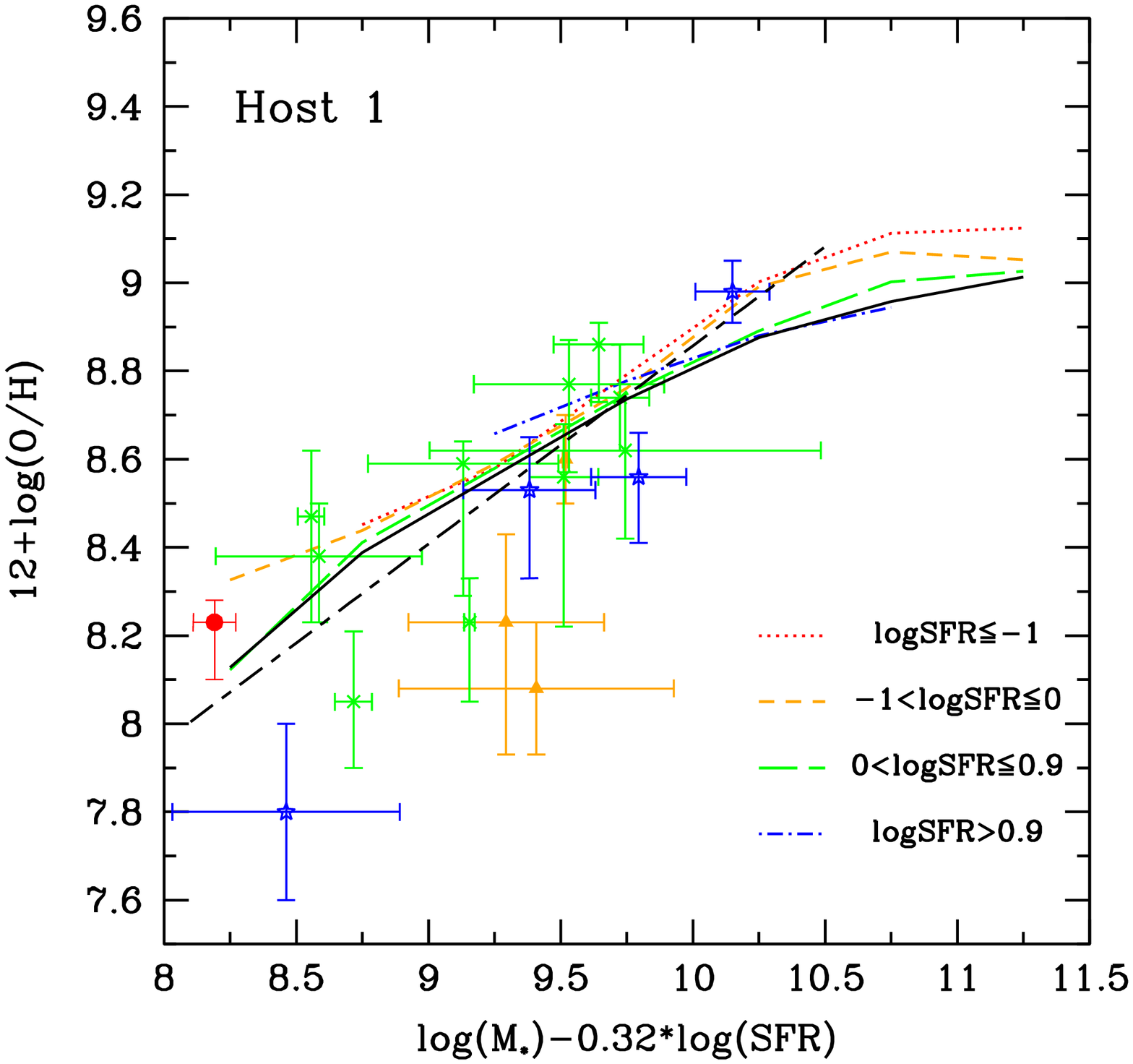}
 \includegraphics[width=7cm] {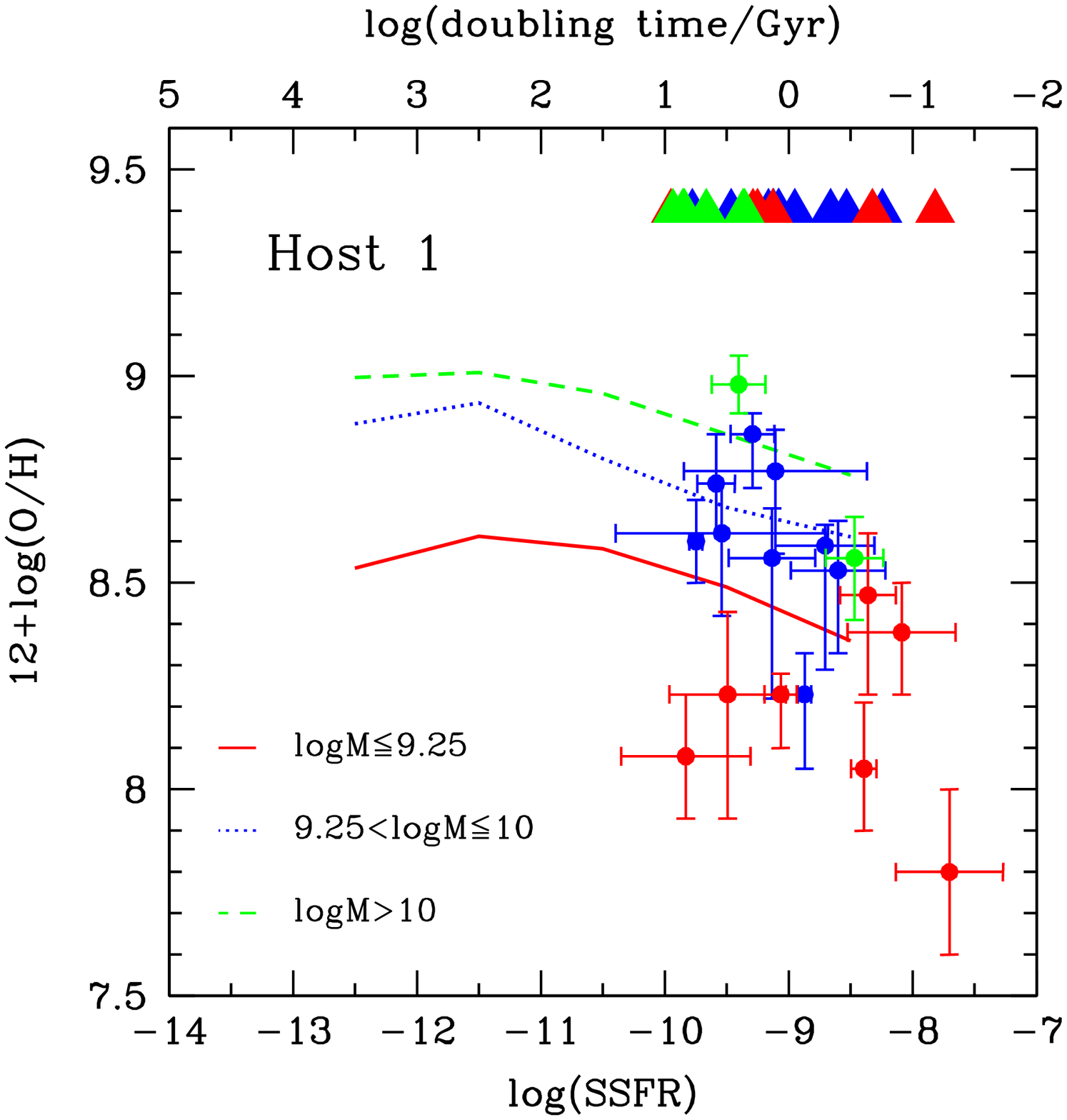}
\includegraphics[width=6.5cm] {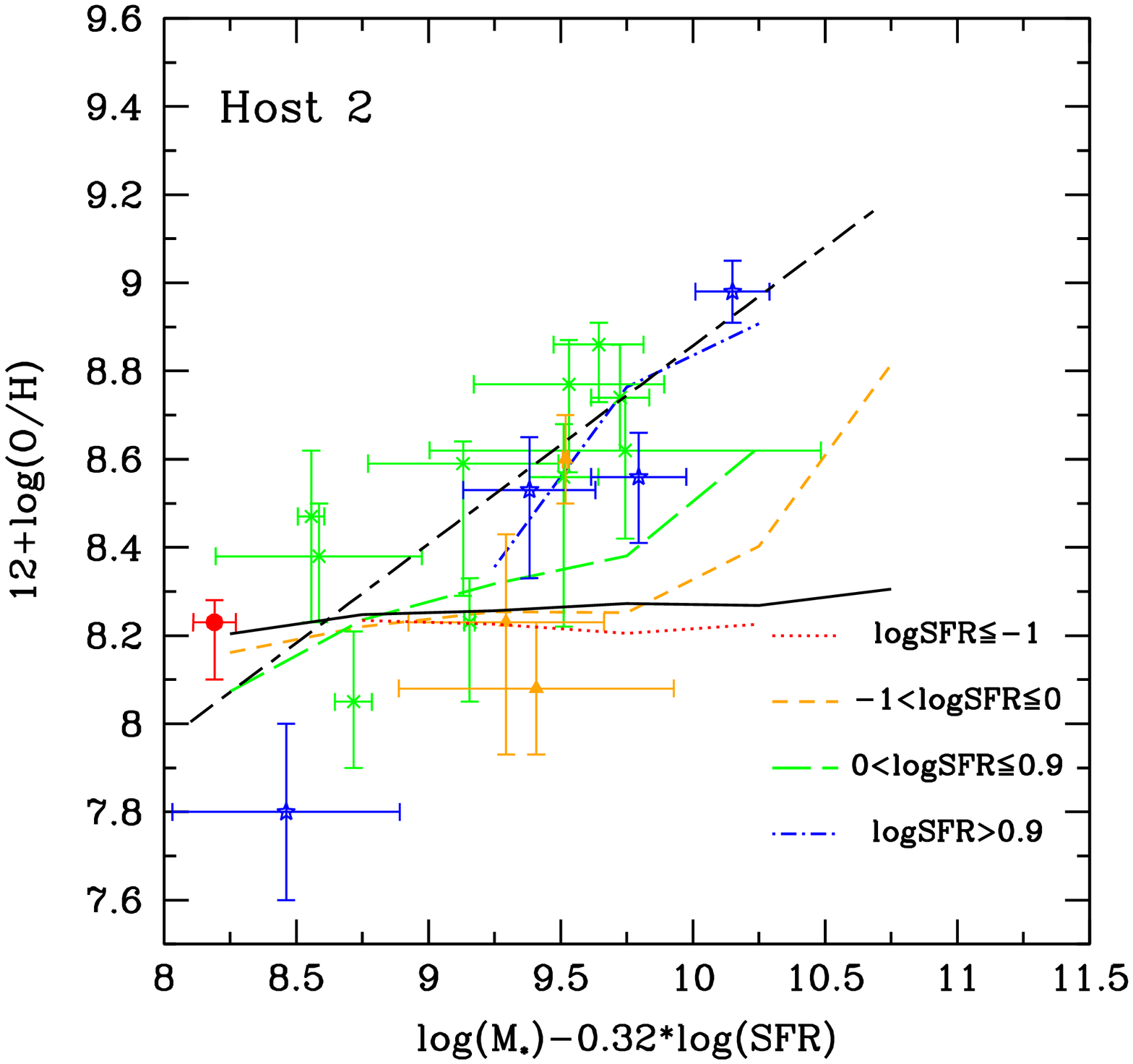}
 \includegraphics[width=7cm] {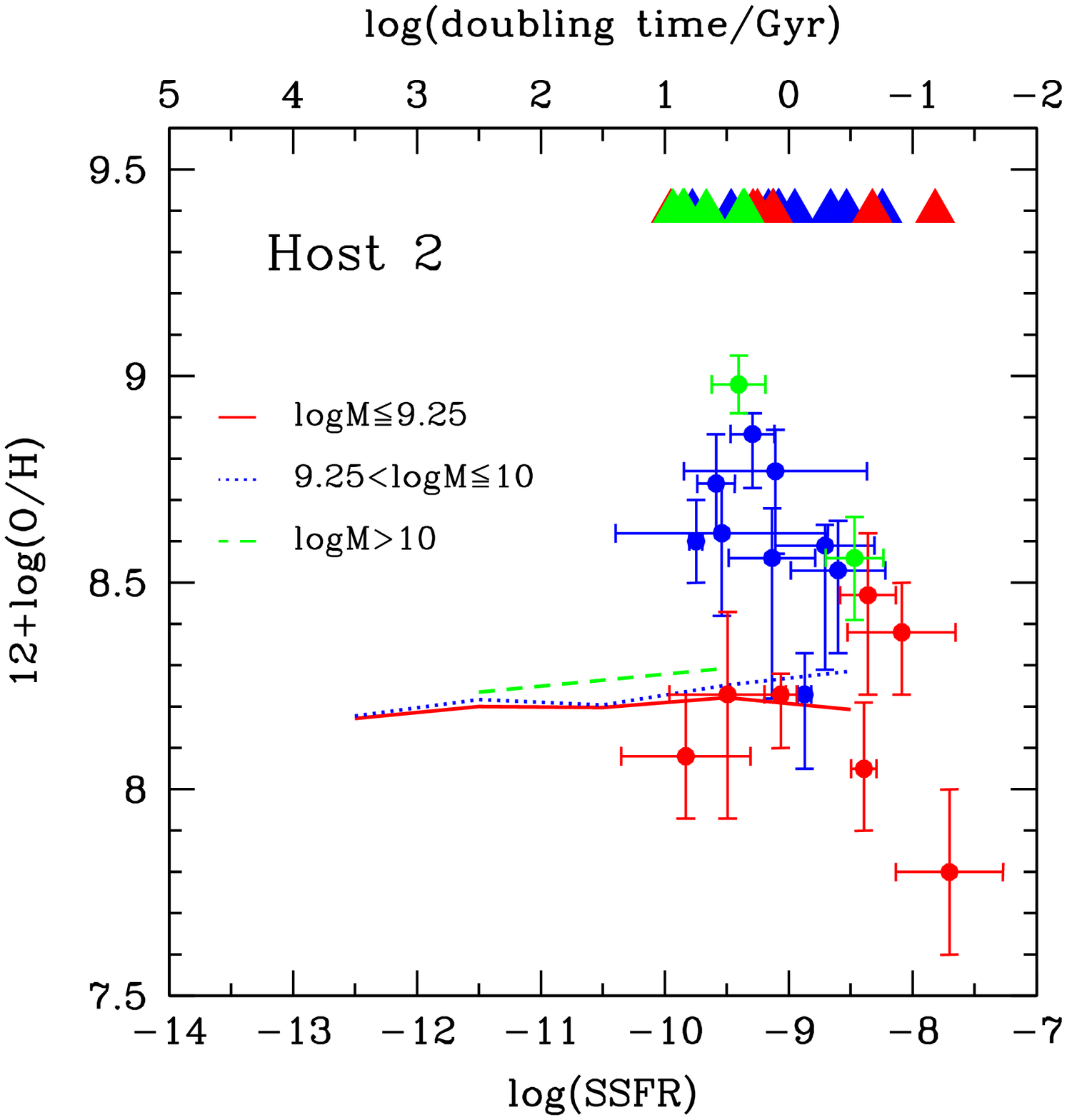}
\includegraphics[width=6.5cm] {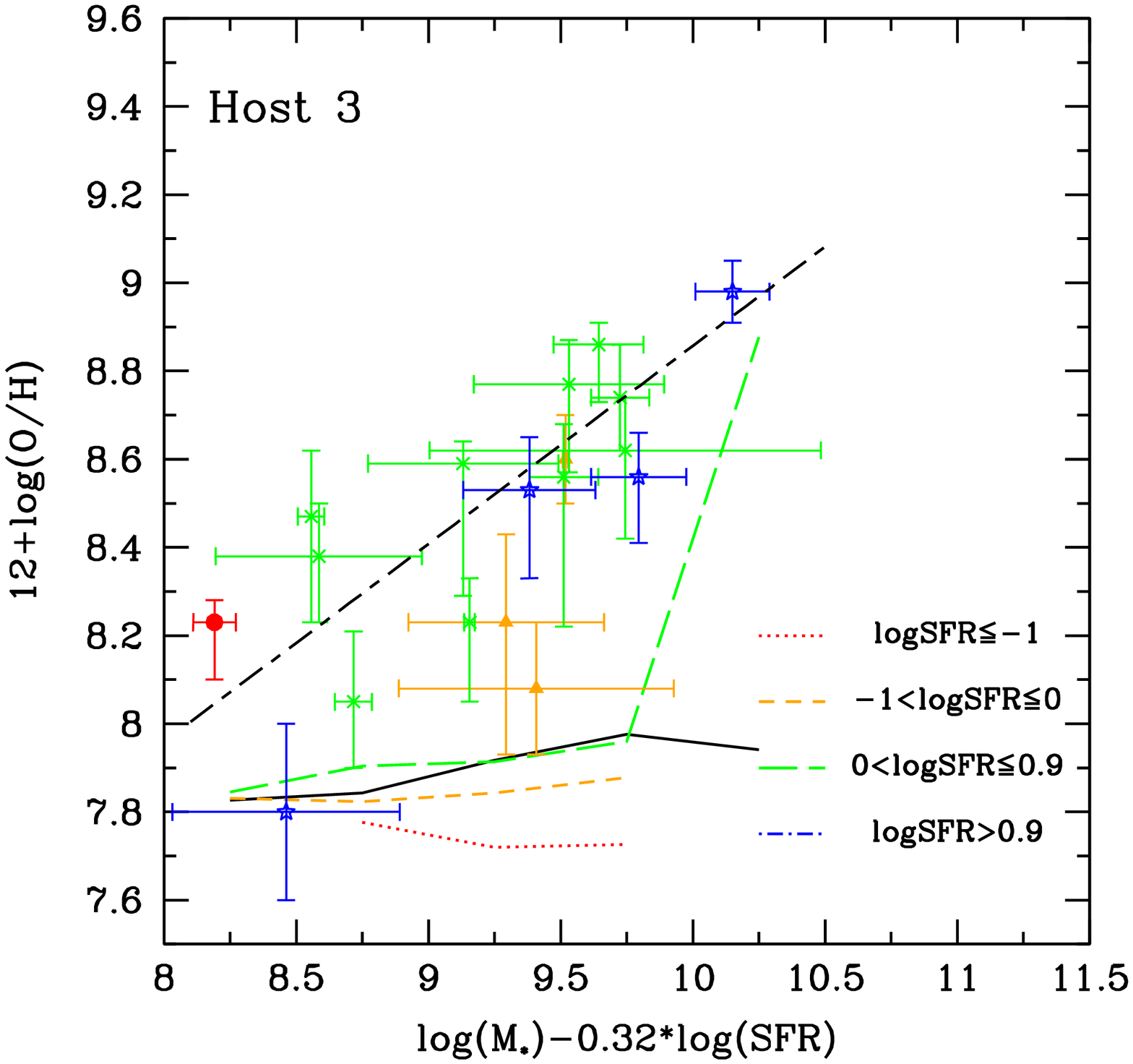}
 \includegraphics[width=7cm] {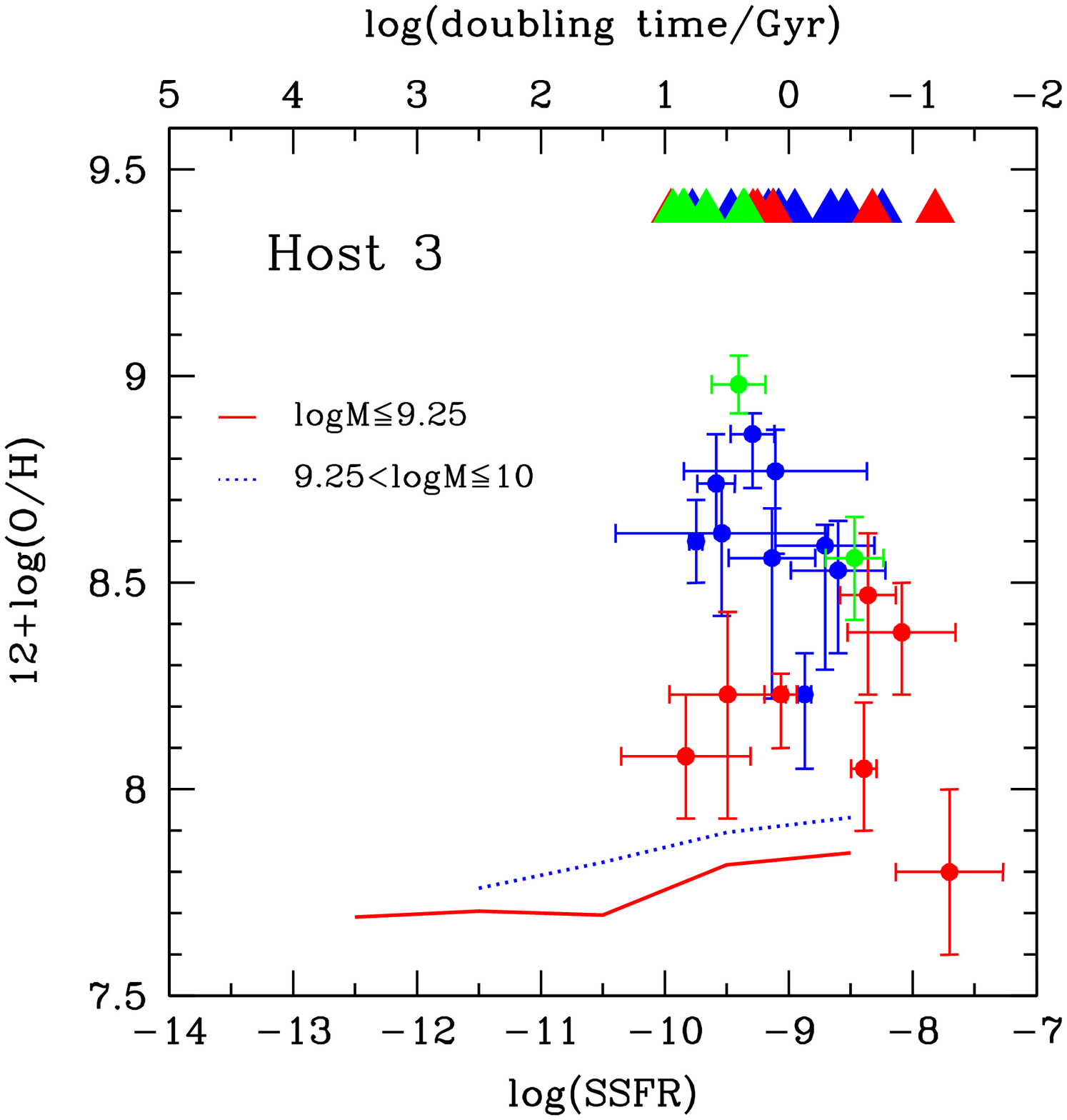}
}
\caption{Left column: $\mu_{0.32}-Z$ plane for the host simulated galaxies  (with $z\sim 0.5$), the colored lines are the median values in different bins of SFR. Data points refer as in Fig.4 to the observed LGRB hosts and black-dashed line is their best fit. Finally black-solid line is the FMR Mannucci et al. 2010a.
Right column: 
SSFR (in units of yr$^{-1}$) versus  metallicity $Z$. Points are the data as in the right panels. Triangles at the top of the figure
refer to 17 LGRBs for which the metallicity is not measured yet. The lines are the median values for HOST1-2-3 in different bin of $M_*$ .}
\label{fig2}
\end{figure*}

In this Appendix, we illustrate further findings obtained from the analysis of HOST1-2-3
in selected  SFR and $M_*$ intervals.
Figure \ref{fig2} show in the $\mu_{0.32}-Z$ and SSFR$-Z$ plane
the relations for our subsamples of host galaxies.
The contribution of hosts with different SFRs are shown with different
colors. The same color coding is adopted for the observed data. Finally, the
short-long dashed line is the best-fit to the observed data obtained by
\cite{Mannucci2010c}. 
The FMR is well reproduced by all subsamples of galaxies belonging to the HOST1 group, and they are
also in agreement with observations (top panel of Fig.~\ref{fig2}).
For the HOST2 sample (middle panel of Fig.~\ref{fig2}) simulated galaxies 
seem to provide a reasonable 
description of the data; both highly star forming objects (blue-dot-dashed lines and data) 
and less active galaxies (yellow-short-dashed and red-dotted lines) and data are well reproduced.
However this model predicts a systematic shift of the LGRB host population
with $0<\log({\rm SFR})<0.9$ (green-dashed lines) toward low metallicities with respect
to the FMR. This shift is not observed in the present sample of LGRB host
data (green points) that represent the bulk of the LGRB host population in
the observed sample. This is reflected in the behavior of the total
FMR of simulated galaxies (solid line) that falls below the observed FMR of LGRB hosts.
This effect is amplified in the HOST3 sample (bottom panel of Fig.~\ref{fig2}-Left Column), where only a few highly star 
forming galaxies with high metallicities are present. The shift in metallicity
for middle and low star forming objects is more evident with the simulated
curves always below the observed data points. 

Fig.~\ref{fig2} shows the distribution of simulated LGRB hosts at $z\sim 0.5$ in the SSFR$-Z$ (right panels).
The lines refer to the median weighted relations in different stellar mass bins,
the data and lines of the same color refer to the same $M_\star$ range.
Again, we find that HOST1 provides a good description of the data.
In particular,  HOST1 galaxies are found to have decreasing metallicities for 
increasing SSFRs and decreasing stellar masses similarly to the observed 
trends. This is a direct consequence of the fact that HOST1 follow the
FMR. Galaxies with higher SFR and 
lower stellar masses have lower metallicities; also higher stellar masses
corresponds higher metallicities. 

In addition, as shown in Fig.~\ref{fig2}, HOST2 and HOST3 sample provide a poor description of the distribution of 
observed LGRB hosts in the SSFR$-Z$ plane. In these cases, the
simulated galaxies show a flat or inverted behavior with respect to the data.
Indeed, the metallicities of simulated hosts are found to increase with 
SSFR. This fact provide further evidence that models requiring a metallicity
threshold for LGRB formation seems at odd with the properties of the observed LGRB host sample.

\end{appendix}

\label{lastpage}

\end{document}